\documentclass[floatfix,a4paper]{article}
\pdfoutput=1
\usepackage{graphicx}
\usepackage{authblk}
\usepackage{amsmath}
\usepackage{amssymb}
\usepackage{xcolor}
\usepackage[top=1in, bottom=1.25in, left=1.25in, right=1.25in]{geometry}
\usepackage{setspace}
\title{\textbf{Dynamic capillary assembly of colloids at interfaces with 10,000$g$ accelerations}}
\author[]{Axel Huerre}
\author[]{Marco De Corato}
\author[]{Valeria Garbin*}
\affil[]{Department of Chemical Engineering, Imperial College London, \\ London SW7 2AZ, U.K. \\*v.garbin@imperial.ac.uk}
\date{}
\begin{document}
\maketitle

\newcommand{\vg}[1]{{\color{magenta} #1}} 
\newcommand{\ah}[1]{{\color{cyan} #1}}
\newcommand{\mdc}[1]{{\color{blue} #1}} 

\textbf{Extreme deformation of soft matter is central to our understanding of the effects of shock, fracture, and phase change in a variety of systems. Yet, despite, the increasing interest in this area, far-from-equilibrium behaviours of soft matter remain challenging to probe. Colloidal suspensions are often used to visualise emergent behaviours in soft matter, as they offer precise control of interparticle interactions, and ease of visualisation by optical microscopy. However, previous studies have been limited to deformations that are orders of magnitude too slow to be representative of extreme deformation. Here we use a two-dimensional model system, a monolayer of colloids confined at a fluid interface, to probe and visualise the evolution of the microstructure during high-rate deformation driven by ultrasound. We observe the emergence of a transient network of strings, and use discrete particle simulations to show that it is caused by a delicate interplay of dynamic capillarity and hydrodynamic interactions between particles oscillating at high-frequency. Remarkably, we find evidence of inertial effects in a colloidal system, caused by accelerations approaching 10,000\textit{g}. These results also suggest that extreme deformation of soft matter offers new opportunities for pattern formation and dynamic self-assembly.}   
\newpage
Particles floating at liquid interfaces are a useful two-dimensional model to visualise the structure and deformation of condensed matter. A fascinating example dates back to L. Bragg, who used rafts of floating bubbles to illustrate grain boundaries and plastic flow in metals \cite{Bragg1942}. Colloidal particles confined at liquid interfaces have been used widely for this purpose, as the interparticle interactions can be finely tuned through electrostatics \cite{Nikolaides_2002,Masschaele2010} and capillarity \cite{Loudet_2005,Cavallaro_2011,Ershov2013}, giving access to a range of two-dimensional condensed phases \cite{Pieranski_1980,Cicuta2003,Madivala_2009,Irvine2012, Keim2014}. The possibility to visualise the structure of the interfacial assembly by optical microscopy has enabled the study of self-healing of curved colloidal crystals \cite{Irvine2012}, of crystal growth \cite{Meng2014} and freezing \cite{guerra_freezing_2018} on curved surfaces, and of dislocations under stress \cite{taccoen2016probing}.  
Beyond their use as two-dimensional model, colloid monolayers at interfaces, and interfacial soft matter in general, play an important role in natural and industrial processes \cite{Sagis2011}. For instance, the mechanical strength imparted to the interface by the colloids enables the formation of bicontinuous emulsions by arrested spinodal decomposition \cite{Herzig2007}, suppression of the coffee-ring effect \cite{Yunker_2011}, and arrested dissolution of bubbles \cite{abkarian2007dissolution,beltramo2017arresting}.

Previous studies of dynamic deformation of colloid monolayers have been limited to relatively low deformation rates \cite{Stancik2002,Cicuta2003,Masschaele2011,Keim2014,Barman_2015,Buttinoni2017} in the range $10^{-2}-1$~s$^{-1}$. However, in realistic conditions, such as in the flow of emulsions and foams, and the evaporation of suspensions, interface deformations can occur on much shorter timescales, driving the system far from equilibrium. In bulk suspensions under flow, colloids form out-of-equilibrium structures stemming from the interplay of interparticle and hydrodynamic interactions \cite{Cheng2012}. Similarly, hydrodynamic interactions between colloids at interfaces can be expected to affect their assembly upon dynamic interface deformation. In more extreme conditions, phenomena that are usually not observed in a colloidal system can become important, for instance as elastic collisions during shock propagation \cite{Buttinoni2017}. Furthermore, some of the phenomena observed for interface deformations of large amplitude, or at high rate, have no counterpart in three-dimensions: when compressed beyond hexagonal close packing, a monolayer of colloids at an interface can buckle out of plane \cite{Pitois2015,Gu_2017} or expel colloids in the surrounding fluid \cite{Poulichet_2015,Gao2016}. Yet, the behaviour of interfacial soft matter under extreme deformation remains poorly understood due to the experimental challenge of simultaneously imparting high-rate deformation and visualising rearrangements of the microstructure. 

\begin{figure}[!htb]
\includegraphics[width=1\textwidth]{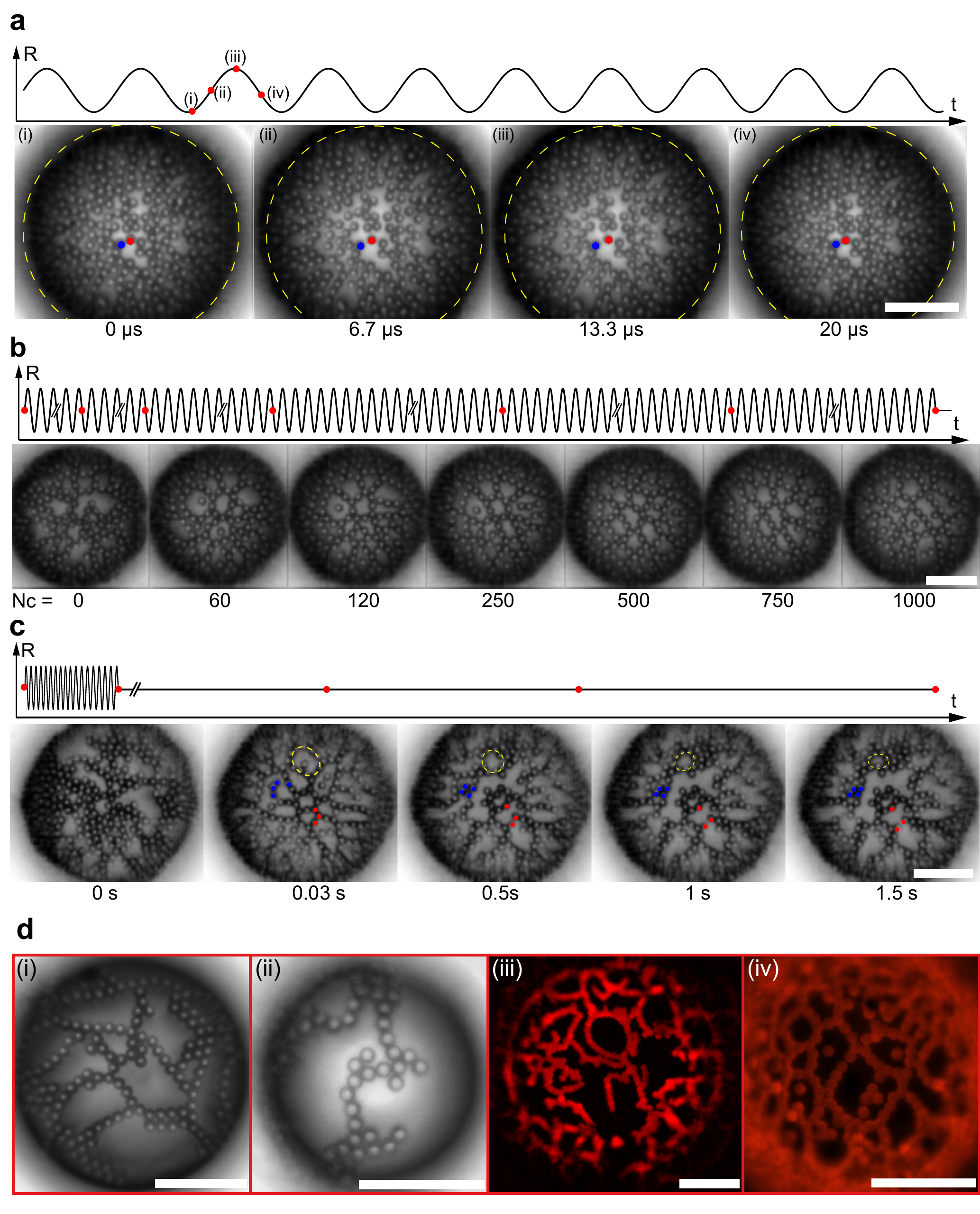}
\caption{\small{ \textbf{Formation of transient microstructure during dynamic deformation of the interface.} \textbf{a}: Time-resolved image sequence of bubble oscillations at 40~kHz ($a=2.5\,\mu$m, $\Phi=0.54$, $R_0=65\,\mu$m, $\Delta R =6\,\mu$m). The colloids marked in red and blue move apart and come back into contact during one oscillation cycle (25~$\mu$s). The dashed circle marks the minimal bubble radius. \textbf{b}: Image sequence of the evolution of the microstructure over 1000 cycles of oscillations ($a=2.5\,\mu$m, $\Phi=0.48$, $R_0=53\,\mu$m, $\Delta R=1.9\,\mu$m). \textbf{c}: Initial aggregated microstructure (0~s), network of strings after 1000 cycles of oscillations (0.03~s) and relaxation of the microstructure after the forcing stops (0.5-1.5~s). Some rearrangements are highlighted by the dashed lines and the color-coded particles ($a=2.5\,\mu$m, $\Phi=0.46$, $R_0=61\,\mu$m, $\Delta R =3\,\mu$m) \textbf{d}: High-resolution bright-field and fluorescence images of networks of strings for different surface coverages: (i) $a=2.5\,\mu$m, $R_0=59\,\mu$m, $\Phi=0.37$; (ii) $a=2.5\,\mu$m, $R=37\,\mu$m, $\Phi=0.2$; (iii) $a=2\,\mu$m, $R_0=86\,\mu$m, $\Phi=0.3$; (iv) $a=2\,\mu$m, $R_0=56\,\mu$m, $\Phi=0.43$. Scale bars: 40~$\mu$m.}}
\label{figure1}
\end{figure}

Here we use acoustic excitation of particle-coated bubbles to explore the far-from-equilibrium phenomena of colloid monolayers at fluid interfaces. Bubbles (equilibrium radius $R_0 \approx 20-100\ \mu$m) were coated with a monolayer of polystyrene spheres (radius  $a\approx 1 - 5\ \mu$m). Electrostatic repulsion between the particles at the interface was completely screened by addition of electrolyte (Methods), so that the initial microstructure of the monolayer at moderate surface coverage was determined by capillary attraction, due to nanoscale undulations of the contact line with an estimated amplitude $Q_2\approx50$~nm \cite{Stamou_2000,Zanini_2017}. This interaction is directional, as can be seen from a decomposition of the interface deformation in two-dimensional multipoles, which shows that contact line undulations result in capillary quadrupoles \cite{Stamou_2000}. Isolated bubbles were driven into periodic compression-expansion by ultrasound at frequency $f=30-50$~kHz in an acoustical-optical setup (Methods), and imaged at 300,000~frames per second. From the high-speed videos, we extracted the evolution of the bubble radius $R(t)$, the maximum oscillation amplitude $\Delta R$, and the trajectories of the particles on the surface of the bubble (Supplementary Information). The surface coverage $\Phi=\frac{N\pi a^2}{A}$, where $N$ is the number of particles in the region of interest, and $A$ is the surface area of the region of interest (Supplementary Information, Figure S1), and the equilibrium bubble radius, $R_0$, were measured from high-resolution still images. \textbf{Figure~\ref{figure1}a} shows a sequence of frames  during one cycle of oscillations of a 65-$\mu$m bubble coated with 5-$\mu$m particles at $\Phi = 0.54 \pm 0.04$. Particles initially at contact are driven apart during bubble expansion, and pushed back into contact during compression, as is clearly seen for the two particles marked in red and blue. The separation of the particles typically occurs in less than 10~$\mu$s.  

A striking change in the microstucture of the monolayer is observed over a few hundreds of cycles of oscillations, as shown in \textbf{Figure~\ref{figure1}b}. Initially the particles, $2.5~\mu$m in radius and with surface coverage $\Phi=0.48 \pm 0.05$, form a disordered, cohesive structure on the interface of a bubble with equilibrium radius $R_0 \approx 53~\mu$m. The arrangement of the particles evolves towards a network of strings in a few hundreds of cycles (Supplementary Movies 1 and 2). The break-up of the initially aggregated structure is made possible by the mechanical energy input provided to the  system during high-rate oscillations. The quadrupolar capillary attraction energy is $E_{\mathrm{quad}} \propto \gamma Q_2^2 \sim 10^4-10^5\ k_{\mathrm{B}} T  $ at contact \cite{Fournier_2002}, resulting in kinetically trapped structures. For comparison, the kinetic energy of a particle in our experimental conditions is $E_{\mathrm{k}}\sim  a^3 \rho f^2 \Delta R^2 \sim 10^3-10^7\ k_{\mathrm{B}} T$.

The network of strings is a transient microstructure that relaxes after the forcing stops. \textbf{Figure~\ref{figure1}c} shows frames of the initial, disordered microstructure before oscillations ($t=0$~s), of the string network formed after 1000 cycles of oscillations ($t=0.03$~s), and of the subsequent relaxation after the forcing has stopped ($t=0.5-1.5$~s). It can be seen that some of the strings break upon relaxation over 1.5 seconds. \textbf{Figure~\ref{figure1}d}(i-ii) shows high-resolution still images of strings formed at low surface coverage, $\Phi \approx 0.37$ in (i) and $\Phi \approx 0.2$ in (ii). Fluorescence images (Methods), shown in \textbf{Figure~\ref{figure1}d}(iii-iv), highlight additional characteristic features of the observed microstructures, such as loops (iii) and lone particles (iv). Loops and lone particles, neither of which are observed at rest, also disappear upon relaxation. 

\begin{figure}[!t]
\includegraphics[width=1\textwidth]{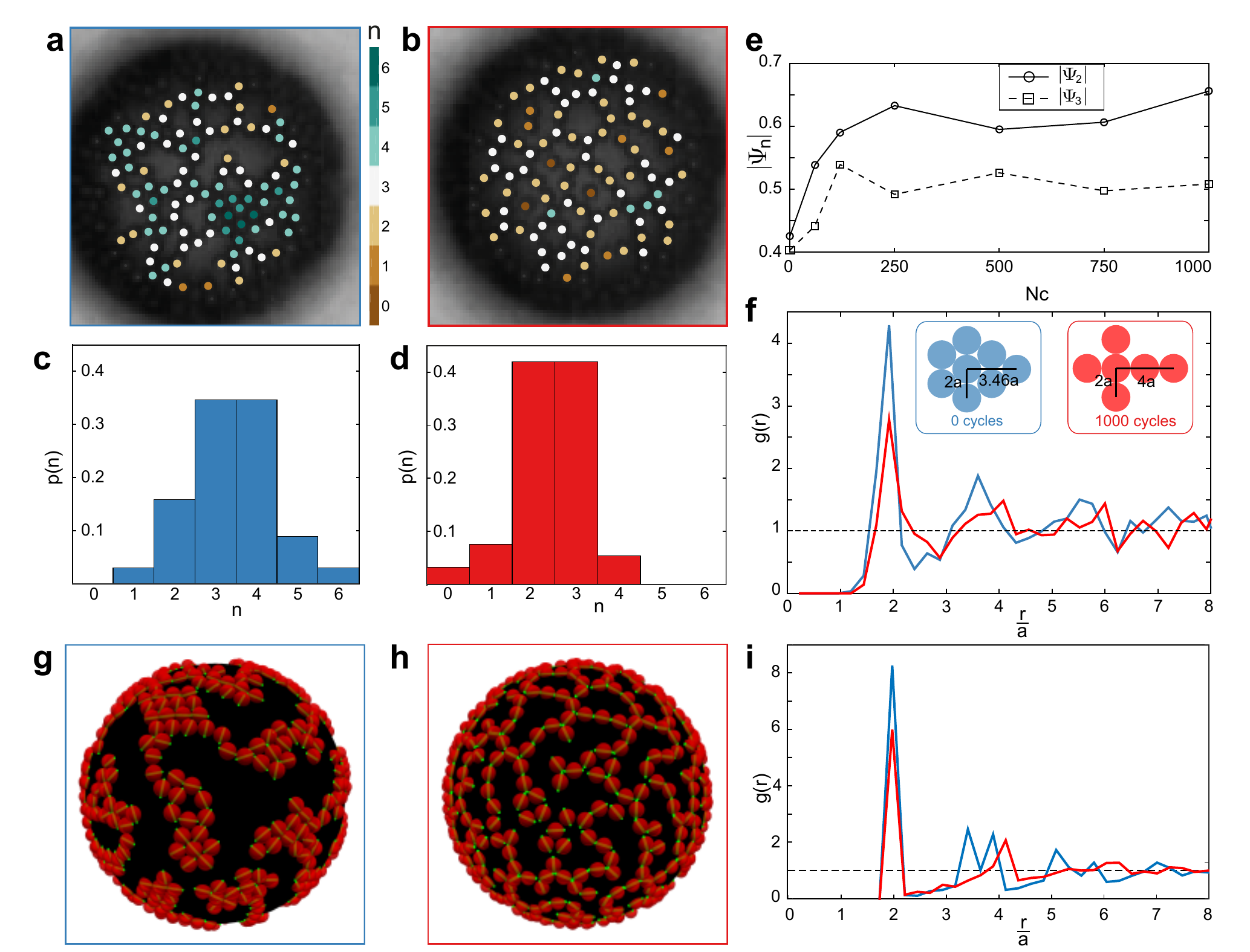}
\caption{\textbf{Quantitative characterisation of the microstructure and comparison with interaction model.}  
 \textbf{a-b}: Images of the initial (a) and final (b) states of the experiment in Figure~\ref{figure1}b, with colour-coding of the particles according to the number of neighbours $n$. The emergence of a network of strings is visually apparent. 
 \textbf{c-d}: Probability $p(n)$ of a particle having $n$ neighbours in the initial (c) and final (d) state. In the final state, $p(n)$ becomes zero for $n=5,6$ and the mean number of neighbours decreases from $\bar n = 3.4$ to $\bar n = 2.4$.
 \textbf{e}: Evolution of the order parameters $|\Psi_2|$ and $|\Psi_3|$ as a function of the number of cycles $N_{\rm{c}}$. The evolution of the microstructure into a network of strings occurs in $\sim 200$ cycles.
 \textbf{f}: Pair-correlation function $g(r)$ as a function of the normalised distance $r/a$ for the initial (blue) and final (red) configurations in experiment. The schematics explain the shift of the second peak upon formation of strings.
 \textbf{g}: Simulation pictures of the initial state ($\Phi = 0.4$). Green lines show the orientation of the quadrupolar deformation. 
 \textbf{h}: Simulation pictures of the final state. The structure obtained is strikingly similar to the experimental one.
 \textbf{i}: Pair-correlation function $g(r)$ as a function of the normalised distance $r/a$ obtained from the initial (blue) and final (red) structures in the simulations. As observed in the experiments, in the final state $g(r)$ has a peak around $r/a =4$, typical of a string network. }\label{figure2}
\end{figure}

A quantitative characterisation of the microstructure shows that the formation of strings is accompanied by a decrease in the number of nearest neighbours, $n$, as can be seen by comparing the initial (\textbf{Fig.~\ref{figure2}a}) and final (\textbf{Fig.~\ref{figure2}b}) states of the experiment of \textbf{Figure~\ref{figure1}b}. In \textbf{Figure~\ref{figure2}a-b}, the particles are colour-coded according to the value of $n$ (Supplementary Information). Initially, particles mainly have 3-5 neighbours. After 1000 cycles of oscillations, particles predominantly have 2 or 3 neighbours. The probability $p(n)$ of a particle having a number $n$ of neighbours is shown in \textbf{Figure~\ref{figure2}c} for the initial state, and in \textbf{Figure~\ref{figure2}d }for the final state. The time evolution of $p(n)$ over 1,000 cycles is shown in Supplementary Information (Figure S3). The statistics show  clearly that particles can have up to 5 or 6 neighbours in the initial state, whereas in the final state $p(n=5)$ and $p(n=6)$ become zero, and there is a sharp increase in $p(n=2)$. Correspondingly, the mean number of neighbours per particle decreases from $\bar{n}=3.4$ to $\bar{n}=2.4$. Not only the number of neighbours is reduced, but the neighbours are aligned, as indicated by the increase of the bond order parameter $|\Psi_2|$ over time, as shown in \textbf{Figure~\ref{figure2}e} (see Supplementary Information). Interestingly, we also observe an increase of the $|\Psi_3|$ order parameter, representing particles having three neighbours organized with a sp$^2$ symmetry. The final structure is therefore formed by particles aligned in strings that are connected respecting a sp$^2$ symmetry. The evolution of the peaks in the pair correlation function, $g(r)$, shows that the nearest neighbours remain at contact ($r=2a$) while the second neighbours move from $r=2\sqrt{3}a$, corresponding to hexagonal close packing, to $r=4a$, corresponding to a chain (\textbf{Fig.~\ref{figure2}f}). All the order parameters considered relax to their initial values after the forcing stops, confirming quantitatively the transient nature of the microstructure (Supplementary Information, Figures S4 and S5). 

\begin{figure}[!htb
]
\includegraphics[width=\textwidth]{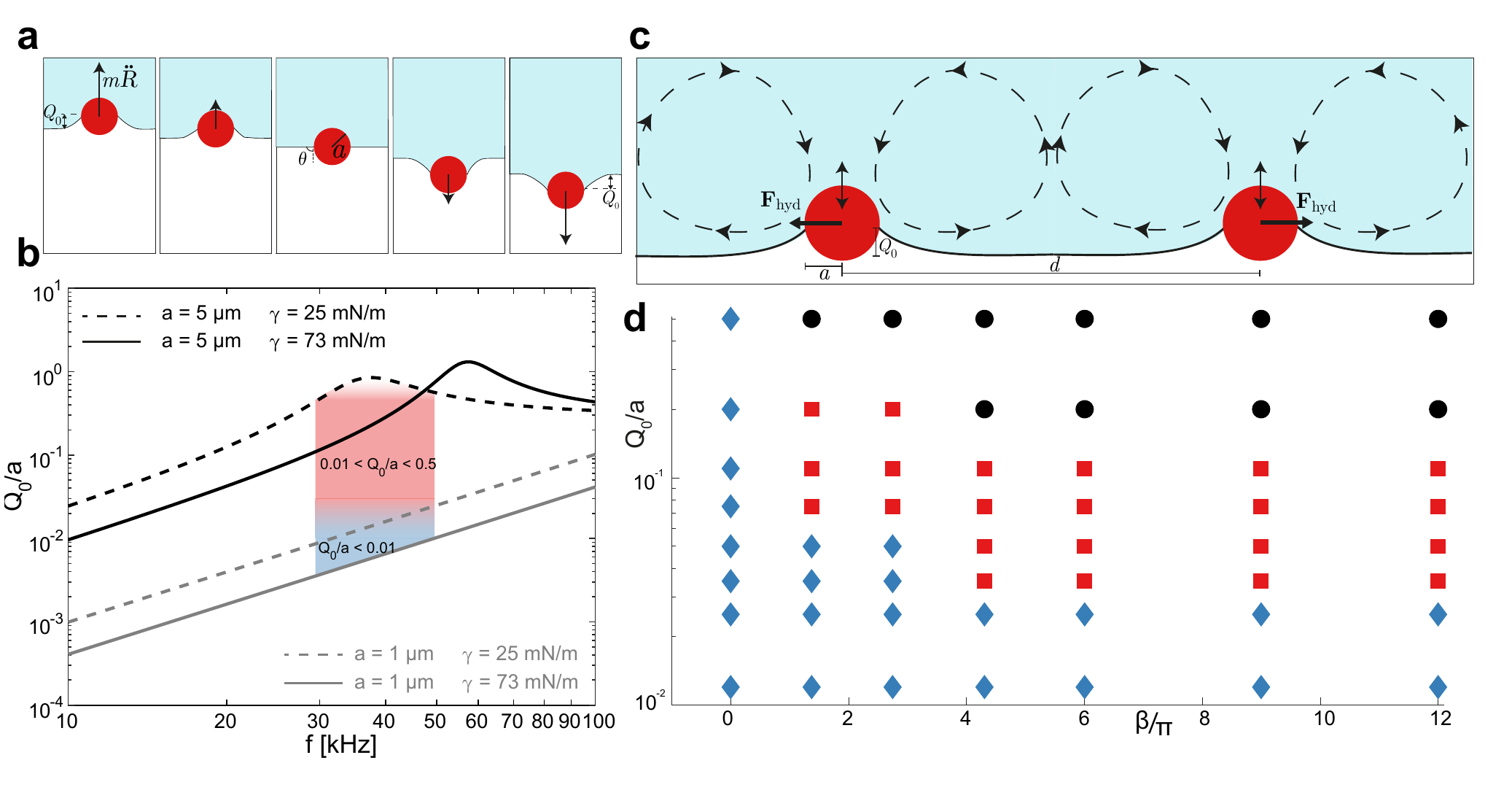}
\caption{\textbf{Capillary and hydrodynamic effects on particles at an oscillating interface.} 
   \textbf{a}: The interface oscillates in the normal direction with acceleration $\ddot{R}$. The inertia of the particle, $m\ddot{R}$, causes a displacement of the particle relative to the interface, resulting in a deformation of the interface of amplitude $Q_0$. 
  \textbf{b}: Amplitude of the monopolar deformation, $Q_0/a$, as a function of the frequency $f=\omega/2\pi$ obtained from a harmonic oscillator model. The particle has mass $m$ and is attached to the interface with a spring of rigidity $\gamma$ (surface tension). Four limiting cases are shown: $a=1\ \mu$m (gray lines) and $a=5\ \mu$m (black lines); $\gamma = 25$~mN/m (dashed lines) and $\gamma = 73$~mN/m (solid lines). The shaded areas correspond to small deformation (blue), and sufficiently large deformation to result in capillary interactions (red).
  \textbf{c}: Two particles oscillating at the interface generate steady recirculating flows leading to repulsive hydrodynamic interactions.
  \textbf{d}: Regime map of the microstructures obtained from particle based simulations as a function of the amplitude of the monopolar deformation, $Q_0/a$, and of the magnitude of hydrodynamic interactions, $\beta$. The final structure after 1000 cycles of oscillations is either an aggregated network (blue diamonds), a network of strings (red squares), or an ordered lattice (black circles).
}\label{figure3}
\end{figure}

We propose that the directional interparticle force leading to the transient formation of strings is due to dynamic capillarity. Hydrodynamic interactions alone, responsible for string formation in the bulk \cite{Cheng2012}, are not sufficient, because they are repulsive for oscillating particles at a fluid interface \cite{singh2009spontaneous}.
Capillary interactions with dipolar symmetry would be sufficient to drive the formation of strings at fluid interfaces \cite{Davies_2014}, but they can be ruled out in our system. Indeed, although dynamic capillary dipoles can be induced by the motion of particles along an interface between two fluids with a large viscosity mismatch \cite{dorr2015driven,dorr2016drag}, in our experiments the viscous stresses due to the lateral motion of the particles are negligible compared to surface tension forces. Another possibility for a directional interaction with dipole-like symmetry is that between a monopole and a quadrupole \cite{Danov_2005} (Supplementary Information, Figure S8). The possibility of a monopolar deformation of the interface in our system is not immediately apparent, because it would only be expected if a body force acts on the particles  \cite{kralchevsky2001particles}. In our experiments, the only body force acting on the particles is gravity, but the Bond number is in the range $Bo=\frac{\Delta\rho g a^2}{\gamma}\approx 10^{-7}-10^{-5}$, hence surface tension forces dominate, and the interface deformation is negligible. We hypothesize that a dynamic monopolar deformation is generated by the motion of the particle relative to the interface during bubble oscillations (\textbf{Fig.~\ref{figure3}a}). The maximum acceleration of the interface is on the order of $\ddot{R}\sim \omega^2 \Delta R  \approx 6000 g$, an extremely large value for a colloidal system, leading to unexpected inertial effects. The Weber number, comparing the inertia of the particle to capillary forces, is $We = \frac{\Delta\rho \omega^2\Delta R  a^2 }{\gamma}\approx 10^{-3}-10^{-1}$ (Supplementary Information) sufficiently large for the resulting interface deformation to drive capillary interactions.

Discrete particle simulations confirm that dynamic capillary interactions between monopole and quadrupole can cause the transient microstructure observed in the experiments. The dynamics of spherical particles confined to the surface of a sphere with time-dependent radius are computed from a force balance on each particle including a simplified model for the capillary and hydrodynamic interactions. The particles are assigned a permanent quadrupolar deformation of amplitude $Q_2$, with an associated orientation vector. The inertia of a particle is assumed to cause a time-dependent monopolar deformation of the interface, $Q(t) = Q_0 \sin (\omega t)$, in phase with the motion of the interface (Supplementary Information, Figure S6). Furthermore, a particle undergoing high-frequency oscillations in a fluid generates a steady recirculating flow, with velocity proportional to $Q_0^2$ \cite{riley1966sphere}.
We model the resulting hydrodynamic repulsion between particles at an interface \cite{singh2009spontaneous} as the viscous drag experienced by a point particle in the streaming flow generated by a neighbouring particle (\textbf{Fig.~\ref{figure3}c}). The pair-wise force is therefore assumed to scale as $F_{\text{hyd}} \approx \beta \, Q_0^2 \, f_{\text{hyd}}(d)$, where $d$ is the interparticle distance, $f_{\text{hyd}}$ is the spatial dependence of the radial streaming velocity field generated by a neighbouring particle (Supplementary Information), and $\beta=6\pi$ for a sphere in a bulk fluid at low Reynolds number. For a particle at an interface, $\beta$ is an unknown function of the Reynolds number, $Re = \dfrac{\rho \, a^2 \omega}{\eta}$, of the capillary number $Ca = \dfrac{\eta a \omega}{\gamma}$, with $\eta$ the viscosity of water, and of the contact angle of the particles. The magnitude of the total interaction force along the line of centers of two particles $i$ and $j$ therefore has the form:
\begin{equation}
F_{\text{int}}^{ij} = Q_0^{2} \, f_{00}(d) \sin{(\omega \, t)}^2 + Q_0 Q_2 \,  f_{02}(d,\varphi_i,\varphi_j)\sin{(\omega \, t)}+Q_2^2 \, f_{22}(d,\varphi_i,\varphi_j) -\beta \, Q_0^2 \, f_{\text{hyd}}(d).
\label{Eq:interaction_model}
\end{equation}
The first term is the interaction between dynamic monopoles, with $f_{00} = 1/(2\pi d)$, the second term is the interaction between a dynamic monopole and a permanent quadrupole, and the third term is the interaction between permanent quadrupoles, with $f_{22} = (48 \pi/ d^5) \cos{\left( 2\varphi_i+2\varphi_j \right)}$. The angles $\varphi_i$ and $\varphi_j$ define the orientation of the quadrupoles of particle $i$ and particle $j$ relative to the line of centers. The function $f_{02} = (4 \pi / d^3) \left[\cos{\left(2 \varphi_i \right)}+\cos{\left( 2\varphi_j \right)}\right]$ accounts both for the effect of monopole $i$ on quadrupole $j$, and of monopole $j$ on quadrupole $i$ (details in Supplementary Information).

We first equilibrate the system with only interactions between permanent quadrupoles ($Q_2/a = 1/100$). The resulting aggregated structure (\textbf{Fig.~\ref{figure2}g}) presents rafts of particles with hexagonal packing and herringbone alignment of the orientation vectors \cite{Nierop_2005}. The dynamics are simulated by allowing the radius of the sphere to vary as $R(t)=R_0+\Delta R \sin (\omega t)$, and by including the effect of the dynamic monopole. The amplitude of the oscillatory monopolar deformation is set to $Q_0/a = 7\times10^{-2}$ and the other parameters are in the range of those used in the experiments (Methods).

The  microstructure after 1000 cycles of oscillations, shown in \textbf{Figure~\ref{figure2}h}, is a network of strings with a striking similarity to the experimental results (see Supplementary Movie 5). The evolution of the pair correlation function $g(r)$ confirms quantitatively the formation of strings (\textbf{Fig.~\ref{figure2}i}). In keeping with the experimental results, the first peak (at $r=2a$) decreases in amplitude and the second peak shifts from $r=2\sqrt{3}a$ to $r=4a$, the signature of evolution from hexagonal close-packing to a string of particles. The orientation vectors of particles inside a string are aligned end-to-end. The  evolution of $p(n)$, $\overline{n}$, $|\Psi_2|$ and $|\Psi_3|$ (Supplementary Information, Figure S11) are all in good agreement with the experimental measurements. The spherical shape of the interface allows to detect the effect of the dynamic monopole, which would otherwise be time-reversible on a planar interface undergoing oscillations, because the interparticle distance is not constant during one cycle of oscillations (Supplementary Information).

To verify that the amplitude of the dynamic monopolar deformation that gives agreement with the experiments is physically justified, we use a harmonic oscillator model of a particle attached to an interface by capillary force \cite{pitois2002small} (Supplementary Information). The non-dimensional displacement $x$ of the particle relative to its equilibrium position provides an estimate of the amplitude of the monopolar deformation, $Q_0/a$. The normalised relative displacement $Q_0/a$ as a function of the oscillation frequency $f=\omega/2\pi$ is shown in \textbf{Figure~\ref{figure3}b} for four cases, using limiting values of the mass $m$ and spring constant $\gamma$ corresponding to the experimental parameters (gray lines: $a=1\ \mu$m; black lines: $a=5\ \mu$m; solid lines: $\gamma = 73$~mN/m; dashed lines: $\gamma = 25$~mN/m). In the range of frequencies used in the experiments, highlighted by the shaded area, the monopole amplitude varies from $Q_0/a\sim 10^{-3}$, up to $Q_0/a\sim 1$. In the range $Q_0/a\sim 10^{-2}-10^{-1}$ (red shaded area in \textbf{Fig.~\ref{figure3}b}), the monopole amplitude is sufficiently large to drive capillary interactions between particles, as has been observed in experiments with heavy particles \cite{Vassileva_2005}. The value $Q_0/a = 7\times10^{-2}$ used in the simulations of \textbf{Figure~\ref{figure2}} is therefore justified. For amplitudes smaller than $Q_0/a \sim 10^{-2}$ (blue shaded area in \textbf{Figure~\ref{figure3}b}), the interface deformation is expected to be insufficient to drive significant interactions. For $Q_0/a \sim 1$ and larger, detachment of particles from the interface can be expected \cite{pitois2002small, Gu_2017}.

Discrete particle simulations predict string formation for physically realistic values of the monopole amplitude, $Q_0$, and of the pre-factor of the hydrodynamic interaction, $\beta$, which are the two unknown parameters in Eq.~\ref{Eq:interaction_model}. We performed a parametric study for values of $\beta$ varying over more than one order of magnitude, and values of $Q_0$ varying in the range predicted by the harmonic oscillator model. The results are presented in the regime map in \textbf{Figure~\ref{figure3}d}. The microstructure obtained after 1000 cycles is classified according to the values of $p(n)$ and $g(r)$ (Supplementary Information) as either an aggregated network (blue diamonds), a network of strings (red squares), or an ordered lattice (black circles). The final microstructure is found not to depend very strongly on the parameter $\beta$. However if $\beta=0$ the structure does not evolve from an aggregated network, suggesting that repulsive hydrodynamic interactions are responsible for the break-up of the initially aggregated structure. The emergence of strings depends strongly on $Q_0/a$, which determines the magnitude of the monopole-monopole attractive force, of the directional monopole-quadrupole interaction, and of the repulsive hydrodynamic force (see Eq.~\ref{Eq:interaction_model}). An aggregated network is obtained when dynamic effects are not important ($Q_0/a \lesssim 10^{-2}$), and capillary interactions between permanent quadrupoles dominate, giving a final structure that remains similar to the initial structure.  Formation of strings is predicted for $Q_0/a \sim 10^{-2} - 10^{-1}$, which is indeed the range in which monopolar deformations are sufficiently large to cause capillary interactions \cite{Vassileva_2005}. When the monopole amplitude exceeds a critical value, $Q_0/a \gtrsim 10^{-1}$, the repulsive hydrodynamic interactions are found to dominate, leading to the formation of an ordered lattice. This ordered structure is never observed in the experiments, likely because it is in the range of deformations where particle expulsion occurs. Additional simulation results showing the effect of each type of interaction on the final microstructure are shown in Supplementary Information, Figure S10.

\begin{figure}[!b]
\centering
\includegraphics[width=1\textwidth]{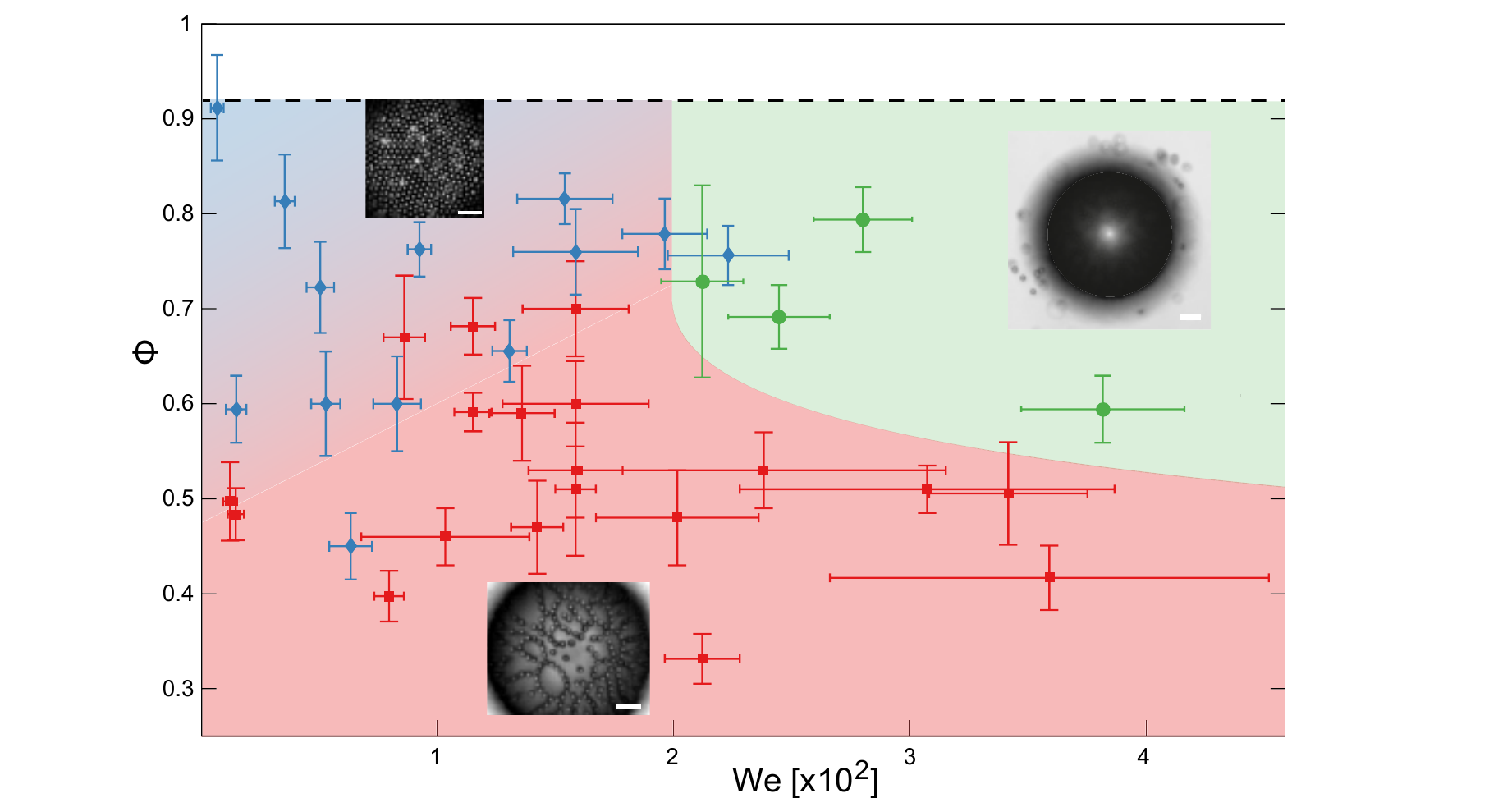}
\caption{\textbf{Experimental regime map: string formation is robust.} The three possible outcomes of an experiment are: formation of strings (red squares); no evolution of the microstructure (blue diamonds); or expulsion of particles (green circles). The shaded areas are a guide for the eye. The regime map shows that the outcome is controlled by the Weber number, $We$, comparing the inertia of a particle with surface tension forces, and the surface coverage $\Phi$. At surface coverages below $\Phi\sim 0.6$, string formation is observed for a broad range of experimental conditions. Error bar calculations are detailed in Supplementary Information. Scale bars: 40~$\mu$m.}
\label{figure4}
\end{figure}

String formation is robust over a broad range of experimental conditions. The two control parameters for the emergence of strings are the surface coverage and the Weber number. The surface coverage determines the extent to which rearrangements of the microstructure are possible. Random close packing of spheres in two dimensions occurs at $\Phi_{\textrm{rcp}}\approx 0.82$, and hexagonal close packing at $\Phi_{\textrm{hcp}}\approx 0.91$ \cite{berryman1983random}. In the experiments, the surface coverage was varied in the range $\Phi\approx 0.33-0.90$ (Supplementary Information, Figure S12). The Weber number, which determines the magnitude of the dynamic monopolar deformation, was varied in the range $We\approx 1 \times 10^{-3} - 5.5 \times 10^{-2}$ by varying the particle size ($2a = 1.8, 3, 5, 10\ \mu$m), the oscillation frequency ($f = 30, 40, 50$~kHz), and the amplitude of oscillations ($\Delta R \approx 0.2-3.5\ \mu$m, with $\Delta R/R_0 \approx 0.003 - 0.13$). We performed over 40 experiments, with three possible outcomes: no evolution of the microstructure, emergence of strings, or expulsion of particles from the interface (see Supplementary Movies 3, 2, and 4 respectively). The results are presented in the ($\Phi$, $We$) plane in \textbf{Figure~\ref{figure4}}. The area highlighted in blue, at high $\Phi$ and low $We$, corresponds to experiments where the structure did not evolve during 1000 cycles of oscillations. In these conditions, rearrangements are limited due to jamming of the monolayer, and the magnitude of the dynamic interactions is insufficient to drive an evolution of the microstructure. The area highlighted in green corresponds to experiments where the amplitude of oscillations is so large that particles are expelled from the interface \cite{Poulichet_2015}. In this regime it is no longer possible to observe structures at constant $\Phi$. Finally, the area highlighted in red corresponds to string formation. Even though this microstructure results from a delicate interplay between hydrodynamics and dynamic capillary interactions, it is consistently observed for a broad range of accessible experimental conditions. 

In summary, we have observed and explained the emergence of a transient microstructure formed by colloids undergoing extreme accelerations at a fluid interface. We have shown that dynamic interface deformation can be used to break kinetically trapped structures, and to impart dynamic capillary interactions. Our results therefore present new opportunities for programmable self-assembly and pattern formation in soft matter at interfaces. Beyond these potential applications, our experimental approach opens the way to future studies of interfacial soft matter far from equilibrium, at deformation rates that are otherwise inaccessible to existing techniques, but relevant to realistic flow and deformation conditions. 

\section*{Methods}

\paragraph*{Particle-coated Bubbles}
Particle-coated bubbles were made by mechanical agitation of an aqueous suspension containing 0.4 \% w/v of colloids, using a vortex mixer. We used hydrophilic, spherical microparticles of nominal diameter 1.8~$\mu$m, 3~$\mu$m, 5~$\mu$m and 10~$\mu$m with sulfate surface groups (IDC surfactant-free latex particles, Life Technologies). For fluorescence imaging we used red fluorescent microparticles, 4 $\mu$m in diameter, sulfate coated (FluoSpheres, Invitrogen). All particles were used as received. Addition of NaCl (VWR Chemicals, AnalaR NORMAPUR, 99.5\%) with 500~mM concentration was necessary to promote particle adsorption to the water-air interface. The resulting Debye length is 0.5 nm. Ultrapure water with resistivity 18.2~M$\Omega$~cm was produced by a Milli-Q filtration system (Millipore).  

\paragraph*{Acoustical-optical setup}
We injected particle-coated bubbles in an observation cell made of a microscope slide and a glass coverslip separated by a 1-mm spacer. 
Prior to every experiment, the cell was rinsed with ethanol and ultrapure water, and dried with compressed air. Small numbers of bubbles were injected in the chamber, and we only observed isolated bubbles (at least 10 diameters away from other bubbles). The observation chamber was placed on an inverted microscope (IX71, Olympus) equipped with 10$\times$ and 20$\times$ objectives. A single-element piezoelectric transducer (SMD50T21F45R, Steminc) with resonance frequency ($45 \pm 3$)~kHz was glued to the glass slide. 
The driving signal was generated by a waveform generator (33220A, Agilent) and amplified by a linear, radio-frequency power amplifier (AG1021, T\& C Power Conversion Inc.). The frequencies used were 30, 40 and 50~kHz, resulting in different bubble oscillation amplitudes. 
Since the wavelength of ultrasound at these frequencies in water is $\lambda >$ 3 cm, the pressure can be considered to be uniform over distances of the order of the bubble size.
The bubbles were driven for 1,000 cycles and the dynamics recorded at 300,000 frames per second using a high speed camera (FASTCAM SA5, Photron). The waveform generator and the high-speed camera are triggered simultaneously using a pulse-delay generator (9200 Sapphire, Quantum Composer). The image resolution at 10$\times$ and 20$\times$ magnification is 2 $\mu$m and 1 $\mu$m, respectively. High-resolution still images in different focal planes were taken before and after excitation at 32$\times$ magnification using a CCD camera (QImaging), resulting in an image resolution of 0.1 $\mu$m. For fluorescence imaging, illumination was provided by a Lumen 200 (Prior Scientific) UV lamp combined with a ET mCH-TR (Chroma) fluorescence cube.

\paragraph*{Particle-based simulations}
The interaction model, introduced in Equation~\ref{Eq:interaction_model}, is presented in Supporting Information. The simulation results in \textbf{Figure~\ref{figure2}} were obtained with the following set of parameters: $\Phi = 0.4$, $\beta = 6\pi$, $\Delta R/R_0 = 0.075$, $R_0/a = 20$, and $Ca = 0.01$.

The model is implemented in a simulation of particles confined to the surface of a sphere with time-dependent radius. Brownian motion is neglected as the timescale for interface deformation is much smaller than the diffusion timescale of the colloids. The evolution of the position $\mathbf{r}_i^*$ and orientation $\mathbf{p}_i$ of particle $i$ on the surface of the bubble is obtained solving the non-dimensional balances of linear and angular momentum:
\begin{equation}\label{force_bal1}
\frac{4 \pi}{3} \, St \, \frac{d \mathbf{v}_i^*} {d t^*} = - 3 \pi \, \mathbf{v}_i^* +\mathbf{F}_S^{*^{i}}+\sum_{j \neq i} \left(\mathbf{F}_{00}^{*^{ij}}+\mathbf{F}_{22}^{*^{ij}}+\mathbf{F}_{02}^{*^{ij}}+\mathbf{F}_{20}^{*^{ij}}+\mathbf{F}_{EV}^{*^{ij}}+ \mathbf{F}_{\text{hyd}}^{*^{ij}}\right) \, ,
\end{equation}
\begin{equation}\label{kinematic_vel1}
\frac{d \mathbf{r}_i^*} {d t^*} =  \mathbf{v}_i^* \, ,
\end{equation}
\begin{equation}\label{angular_momentum_bal1}
\frac{8 \pi}{15} St \frac{d \boldsymbol{\omega}_i^*} {d t^*} = - 4 \pi \, \boldsymbol{\omega}_i^* +\sum_{j\neq i} \left(\mathbf{T}_{22}^{*^{ij}}+\mathbf{T}_{20}^{*^{ij}} \right) \, ,
\end{equation}
\begin{equation}\label{kinematic_rotvel1}
\frac{d \mathbf{p}_i} {d t^*} =  \boldsymbol{\omega}_i^* \times \mathbf{p}_i \, .
\end{equation}
The forces and the torques on the right-hand sides represent the contributions of capillary, hydrodynamic, and excluded-volume interactions between particle $i$ and particle $j$ and of the surface constraining force (Supplementary Information). 
Equations \eqref{force_bal1}-\eqref{kinematic_rotvel1} are solved using a second-order explicit linear multistep method. We used a nondimensional time step $\Delta t^*= 3 \times 10^{-3}$, which has been checked to give convergent results. To avoid expensive computations we enforce a cut-off length of $7 a$ for all the interaction forces and torques between the particles. Independent simulations with larger and smaller cut-off lengths were found to give similar results. The computational time of evaluating the interactions between the particles scales as $N^2$. The computation of the interactions was therefore parallelised over ten or more cores, to speed up the computations. 

\section*{Acknowledgements}
We thank L. Botto and A. Striolo for useful discussions. This work is supported by the European Research Council, Starting Grant No. 639221 (ExtreFlow). 

\section*{Author Contributions}
A.H. and V.G. designed the experiments. A.H. performed the experiments and analysed the data. M.D.C. designed and implemented the simulations. A.H., M.D.C., and V.G. interpreted the results and wrote the manuscript. V.G. designed the research.
\bibliographystyle{naturemag}
\bibliography{bibliography}

\end{document}